# Tuning the porosity of bimetallic nanostructures by a soft templating approach

*Anaïs Lehoux,[1,2] Laurence Ramos,[3,4] Patricia Beaunier,[5] Daniel Bahena Uribe,[6] Philippe Dieudonné,[3,4] Fabrice Audonnet,[7] Arnaud Etcheberry,[8] Miguel José-Yacaman,[6] Hynd Remita[1,2]\**

1- Laboratoire de Chimie Physique, UMR 8000-CNRS, Université Paris-Sud, 91405 Orsay, France, E-mail: hynd.remita@u-psud.fr
2- CNRS, Laboratoire de Chimie Physique, UMR 8000, 91405 Orsay, France
3- Université Montpellier 2, Laboratoire Charles Coulomb UMR 5221, F-34095, Montpellier, France
4- CNRS, Laboratoire Charles Coulomb UMR 5221, F-34095, Montpellier, France
5- Laboratoire de Réactivité de Surface, UPMC Université Paris 6, UMR 7197-CNRS, 3 rue Galilée, 94200 Ivry, France
6- Department of Physics & Astronomy, The University of Texas at San Antonio, One UTSA Circle, San Antonio, TX 78249, USA
7- Institut Pascal, UMR 6602-CNRS, Université Blaise Pascal, F-63177 Aubière Cedex, France
8- ILV-Institut Lavoisier de Versailles, 45 rue des Etats-Unis, UMR 8180-CNRS, Université Versailles Saint-Quentin-en-Yvelines, 78035 Versailles Cedex, France

**Keywords:** Bimetallic Nanoparticles, Palladium-Platinum Alloy, 3D-Porous Nanostructures, Soft Templates, Radiolysis.

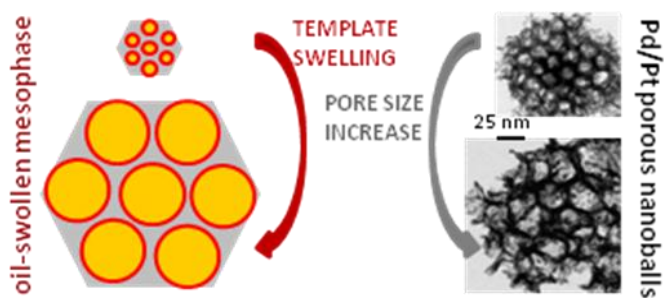

## Abstract

We use hexagonal mesophases made of oil-swollen surfactant-stabilized tubes arranged on a triangular lattice in water and doped with metallic salts as templates for the radiolytic synthesis of nanostructures. The nanostructures formed in this type of soft matrix are bimetallic palladium-platinum porous nanoballs composed of 3D-connected nanowires, of typical thickness 2.5 nm, forming hexagonal cells. We demonstrate using electron microscopy and small-angle X-ray scattering that the pore size of the nanoballs is directly determined by the diameter of the oil tube of the doped mesophases, which we have varied in a controlled fashion from 10 to 55 nm. Bimetallic nanostructures comprising various proportions of palladium and platinum can be synthesized. Their alloy structure was evidenced by X-ray photoelectron spectroscopy, energy-dispersive X-ray spectroscopy, and high-angular dark field scanning transmission electron microscopy experiments. Our templating approach allows therefore the synthesis of bimetallic nanoballs of tunable porosity and composition.

**Introduction**

Mesoporous precious metals, especially platinum and palladium are of great interest because of their potential applications in different fields such as catalysis,[1] electrocatalysis and sensing.[2] Indeed, platinum and palladium play a crucial role in catalysis and are involved in various reactions such as hydrogenation and dehydrogenation reactions[3] and CO oxidation in catalytic convertor.[4] Pt- and Pd-based nanoparticles are also very promising electrocatalysts for fuel cells applications.[5] In addition, Pd-based materials have been shown to absorb a large quantity of hydrogen and to display remarkable performances in $H_2$ storage and sensing.[6-10] On the other hand, because of the scarcity and the high cost of Pt, optimizing both the size and shape of Pt-based nanostructures is necessary to reduce Pt loading,[11-13] while keeping the same catalytic or electrocatalytic performance of the materials. To reach this goal, two ways can be explored: increasing the surface over volume ratio by a size reduction and improving the activity by creating specific shapes that will expose preferential crystallographic planes. In this optics, porous nanostructures are very promising as they are expected to exhibit enhanced catalytic activities since they can supply different adsorption sites for the reactions involving two or more reactants and they may allow unlimited transport of molecules of the reactive medium.

Bimetallic nanoparticles (NPs) have attracted increasing attention because they are expected to exhibit new solid-state properties, which differ from their monometallic counterparts.[14] Therefore, they have potential applications in various fields such as catalysis, electrocatalysis, sensing, magnetic storage… Bimetallic NPs often exhibit enhanced catalytic performances in terms of activity, selectivity and stability, compared to the separate components.[14-16] Indeed, PdPt alloy NPs are promising catalysts for hydrogenation of aromatic hydrocarbons and for electrooxidation of small organic molecules with high sulfur [17, 18] and CO tolerance.[19, 20] Once deposited on $TiO_2$ alloyed PtPd NPs exhibit high photocatalytic activity for elimination of CO and volatile organic compounds in the presence of humidity.[21] Recent studies indicate also that Pt-Pd bimetallic nanostructures present great opportunities in developing novel electrocatalysts for polymer electrolyte membrane fuel cells. Binary Pt/Pd nanoparticles obtained by localized overgrowth of Pd on cubic Pt seeds exhibit higher activity for formic acid oxidation as compared to Pt nanocubes, Pd preventing poisoning.[22] In addition, Pt-Pd bimetallic nanostructures were found to be very promising for oxygen reduction[23-25] and Pt-Pd alloys nanoparticles and Pt-on-Pd bimetallic nanodendrites synthesized on graphene nanosheets exhibit high electrocatalytic activity for methanol oxidation.[26, 27] Recently, it has also been shown that the Pt-Pd nanocages exhibit both enhanced activity and selectivity for the preferential oxidation of CO in excess of hydrogen than those of Pd nanocubes and commercial Pt/C thanks to the alloy composition and hollow structure.[28]

Hence, because of the large variety of potential applications, synthesis of Pt- and Pd- based nanostructures and in particular bimetallic Pt-Pd nanoparticles with controlled sizes, compositions, shapes and porosity is a very active research area.[25, 29-31]

Mesophases resulting from surfactant self-assembly are versatile templates for generating one-, two- or three-dimensional nanostructures in relatively large quantities which can be extracted easily by the dissolution of the mesophase. Soft templating generally offers a high degree of control over the pore size as well as microstructure periodicity. Since the pioneer work of Attard and co-workers, who prepared mesoporous Pt from lyotropic liquid-crystal templates,[32, 33] various syntheses of metal nanostructures using soft template mesophases have been reported.[34-42] Most of the syntheses have resulted so far in materials with unidimensional porosity, such as arrays of tubes.

In the past years, we have used as templates a versatile class of mesophases composed of oil-swollen tubes, of tunable diameter, which are stabilized by a monolayer of surfactant and co-surfactant molecules, and packed on a triangular lattice in a continuous aqueous phase.[43, 44] We have shown that swollen hexagonal mesophases can be doped with various compounds and used as nanoreactors to synthesize nanostructured materials (metal, polymer or oxides) both in the aqueous and in the oil phases.[45-47] Recently, Pd nanowires,[48] Pd hexagonal nanosheets[49] and porous Pt,[50] Pd[51] and Pd-Au[52] nanostructures have been synthesized in these soft moulds.

So far, very few examples have been reported for the preparation of three-dimensional porous bimetallic nanostructures.[38], [52, 53] Here, we describe the soft template radiolytic synthesis of three-dimensional porous bimetallic Pd-Pt nanostructures made of interconnected nanowires forming hexagonal cells. We demonstrate that a direct templating effect of the hexagonal mesophases, whose characteristic sizes can be tuned by a factor 5, controls the pore size.

## RESULTS AND DISCUSSION

**Nanoballs with tunable porosity**

The hexagonal mesophases are made of a mixture of cetyltrimethylammonium bromide (CTAB), as surfactant, pentanol as cosurfactant, cyclohexane, water and salt (NaCl and/or metallic salts). Based on previous results[43, 44], in order to vary the size of the oil-swollen tubes while remaining in a one-phase region of the phase-diagram, one has to vary in a concomitant way the amount of oil, cyclohexane, and the salt concentration. In our experiments, the salt concentration is varied between 0.1 M and 1 M, and the corresponding swelling ratio defined as the volume ratio of cyclohexane over water, O/W, is varied between 1.5 and 6.7. As controlled experiments, we checked by small-angle X-ray scattering (SAXS) that the mesophases prepared with only NaCl as salts and whose composition are given in **table 1** are hexagonal mesophases. The lattice parameter ($d_c$), diameter of the oil-swollen tubes (D) and thickness (e) of the water channel between adjacent tubes of these hexagonal mesophases are also given in **table 1**. We found a monotonic and linear increase of the tube diameter as O/W increases (**table 1** and **Figure S1 in SI**), from 13.2 to 60.4 nm, whereas the water channel is slightly decreasing (from 2.8 to 1.3 nm). An extrapolation of the diameter towards 0 swelling ratio yields a value of 2.4 nm, which is comparable to twice the thickness of the alkyl chain of the surfactant in all-trans position in solution (about 1.7 nm[54]), intercrossed on a third of their length.[55] Note that, a two-fold increase of maximum swelling and maximum tubes diameter is obtained here as compared to previously published results[43, 44] (maximum swelling ratio, respectively tube diameter were about 3.8 resp. 34 nm, as compared to 6.7 and 60 nm obtained in the present study).

In **Figure 1a**, we show the SAXS patterns of several samples with various swelling ratio O/W and doped with 0.1 M of metallic salts (Pd/Pt = 1:1). All samples were hexagonal, as checked from the occurrence and the relative positions of the several Bragg diffraction peaks. Moreover, we find that the peaks shifted towards lower wave vectors as the swelling ratio increases, indicating an increase of the characteristic sizes of the mesophases. From the peak positions and samples composition, we have calculated (as described in the Materials and methods section), the diameter of the oil-swollen tubes. Results are reported in **Figure 2** as a function of the swelling ratio. Similarly to what has been measured with NaCl, the diameter varies linearly with the swelling ratio, and increases from 15 to 55 nm, when O/W increases from 1.5 to 6.1. Comparable values are obtained with NaCl and with samples doped with 0.1 M metallic salts **(Figure S1)**.

The Pd-Pt doped mesophases were used as a nanoreactor to synthesize Pd-Pt nanostructures. The samples were exposed to γ-irradiation at a relatively low dose rate for a slow reduction. The hydrated electrons and the reducing radicals produced during the radiolysis of water induce homogeneous reduction in the confined aqueous phase. The pentanol, used as cosurfactant, contributes to $Pd^{II}$ and $Pt^{II}$ reduction on the radiolysis-induced seeds. Compared to chemical reducing processes that follow a diffusion front, radiolysis presents the advantage to induce a homogeneous nucleation and growth in the whole volume.[56] Our experiments show that the mesophase nanostructure is preserved upon irradiation of the metallic salts. As an illustration, we provide in **Figure S2** the SAXS patterns obtained before and after irradiation, of a sample with a swelling ratio O/W = 2.5, doped with 0.1 M metallic salt (and comprising also NaCl at 0.2 M). **Figure S2** shows that the two patterns are qualitatively similar, with the occurrence of three Bragg diffraction peaks of equivalent width and whose positions confirm a hexagonal symmetry, with a slight shift of the peaks towards higher wave vectors for the SAXS pattern of the sample after irradiation. The shift towards high $q$ is systematically observed for samples after irradiation and is the signature of a slight contraction (at most 20%) of the characteristic size of the mesophase upon irradiation.

Transmission electron microscopy (TEM) observations revealed that the materials synthesized in the hexagonal mesophases are porous nanoballs of diameter in the range 100 nm - 280 nm diameter **(Figure 3 and S3).** Although very noisy, experimental data show that the average size of the nanoballs increases steadily from 100 to 280 nm as the swelling ratio of the templating mesophases increases from 1.5 to 6, as shown in **Figure S4**. These porous structures are constituted of parallel or three-dimensionally interconnected nanowires forming the walls of hexagonal cells. The typical diameter of the nanowires is 2.5 nm, a numerical value in remarkable agreement with the thickness of the water channels between adjacent tubes in the template mesophase (between 1.3 and 2.8 nm). The bimetallic composition of the nanoballs was confirmed by energy-dispersive X-ray spectroscopy (EDS) analysis (**Figure S5**).

High resolution transmission electron microscopy (HRTEM) images revealed the detailed structure of the nanoballs and clearly show (**Figures 4a and S6**) several hexagonal cells delimited by polycrystalline nanowires. The fast Fourier transform (FFT) presented two reciprocal distances $d_1$= 0.226 nm and $d_2$= 0.196 nm corresponding to the (111) and (200) crystal planes of Pd-Pt nanostructure. HRTEM image of peripheral nanowires (**Figure S6**) showed that they are formed of interconnected nanocrystallized domains showing the (111) and (200) lattice fringes. The formation of twin boundaries and stacking faults are also seen along all the nanowires of the hexagonal cells (**Figure 4a1 and Figure 4a2**).

Irradiations carried out on doped mesophases with different swelling ratios and hence different oil tube diameters produced qualitatively similar porous nanoballs (**Figure 3).** (Metallic nanostructures synthesized in hexagonal mesophases with other swelling ratios are provided in **Figure S7**). Interestingly enough, TEM observations show that the average pore size of the Pd-Pt nanostructures increases with the diameter of the oil cylinder of the soft template from 10 to 52 nm. In **Figure 2,** we show the dependence of the pore size with the swelling ratio, O/W, of the template hexagonal phase. Remarkably, this set of experimental data collapses perfectly with the evolution of the tubes diameter of the hexagonal phase, plotted as well in **Figure 2**, thus demonstrating a direct templating effect of the mesophase.

Because results on TEM rely on the analysis of a relatively small number of images, an independent measurement of the characteristic pore size of the nanoballs, averaged over a large assembly of nanoballs, has been performed using small- and wide-angle X-ray scattering. Given the range of wave vectors, $q$, experimentally accessible (0.2 – 20 nm$^{-1}$), the internal structure of the nanoballs was probed in these experiments. We show in the **inset of Figure 1b** the X-ray intensity, $I$, scattered by an assembly of bimetallic nanoballs synthesized in a mesophase with a swelling ratio, O/W=1.5. The scattered intensity is characterized by a weak peak at small wave vectors, followed by a power law decrease of $I$ vs $q$, with an exponent between -2 and -3, followed at even lower $q$ by a rather featureless decrease of the scattered intensity. The peak, at $q^* \cong 0.4$ nm$^{-1}$, indicates a characteristic size, $2\pi/q^* \cong 15.6$ nm, which can be identified as the nanoballs pore size (for the same sample, analysis of TEM images gives a size of 14.3 nm). The power law decrease reflects the complex internal structure of the porous nanoballs; the cross-over from this regime to the featureless regime occurs at a wave vector of $q^{**} \cong 1.4$ nm$^{-1}$. It corresponds to a characteristic size, $2\pi/q^{**} \cong 4.5$ nm, of the order of the thickness of the metallic wires of the porous nanoballs, although slightly larger. Nanoballs prepared in mesophases with various swelling ratio were measured. We find that the position of the peak at low wave-vectors, $q^*$, is shifted towards low $q$s when the swelling of the templating hexagonal mesophases increases, in agreement with an increase of the pore size as measured by TEM (**Figure 1b).** The dependence with the swelling ratio of the pore size, $2\pi/q^*$, is reported in **Figure 2.** The numerical values show a remarkable agreement with the values measured from TEM images, and the tubes diameter of the hexagonal mesophase, thus confirming the direct templating effect of the mesophase.

**Nanostructure growth**

In previous studies conducted on pure Pd and Pt based mesophases, we have shown that slow reduction was necessary to obtain these porous 3D-nanostructures.[50] To further understand the growth process for the porous 3D-nanostructures, samples at different doses were observed by TEM. After 5 hours of irradiation (**Figure 5a**), the formation of individual nanoparticles smaller than 2 nm were observed. After 12 hours (**Figure 5b**) nanostructures of about 50 nm formed by nanowires were observed. These "embryonic" dendrites grow with irradiation time and after 24 h of irradiation (**Figure 5c**), the 3D-dendrites have grown much larger (about 80 nm) and hexagonal cells were observed. After 48 h irradiation, larger porous nanoballs (of about 100-120 nm for O/W = 1.5) were formed by a more dense and well organized network of nanowires: parallel or 3D-interconnected nanowires forming hexagonal cells (**Figure 5d**).

In order to study the effect of the surfactant on the nanoparticle growth, experiments have also been conducted with mesophases based on other cationic surfactants cetyltrimethylammonium chloride CTAC and cetylpyridinium bromide CPBr. While very ordered 3D nanostructures (made of parallel nanowires or connected ones forming hexagonal cells) are obtained with CTAB. Nanostructures made of parallel nanowires (of 2.4 nm) are also formed with CTAC, but in this case the nanostructures are much less ordered and no hexagonal cells were observed (see **Fig.S8 b and c** in supplementary information). In the case of CPBr, the obtained nanostructures were spherical aggregates made by nanoparticles of 3-4 nm. These experiments attest the role of $CTA^+$ and $Br^-$ in promoting the 1D- to 3D-growth. Many studies have indeed shown that halides, and particularly $I^-$ and $Br^-$ adsorb tightly on single-crystalline Pd surfaces[57, 58] and that $Br^-$ adsorbs preferentially on (100) facets. In this respect, the cetyltrimethylammonium bromide ($CTA^+$, $Br^-$) surfactant is considered to play a key role as a "complexing agent" and/or as a structure-directing agent.

In another step, we investigated the role of the reducing species on the nanoparticle growth. Irradiation of the water phase induces formation of solvated electrons $e^-_{aq}$ ($E^0(H_2O/e^-_{aq})$ = -2.87 $V_{NHE}$),[59] and of $H^\cdot$ and $OH^\cdot$ radicals. Because of their reaction with the cosurfactant (pentanol), these radicals are converted into reducing alcohol radicals. Most synthesis were conducted under $N_2$ atmosphere where both solvated electrons and alcohol radicals contribute to the reduction process. Experiments have also been conducted under $N_2O$ atmosphere (solubility in water at 25 °C and 1 atm. = 25 mM)[59] in order to scavenge the solvated electrons:[60-62]

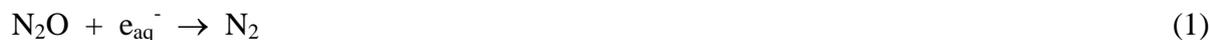
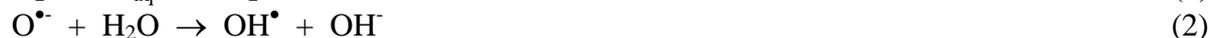

$$N_2O + e_{aq}^- \rightarrow N_2 \qquad (1)$$
$$O^{\bullet-} + H_2O \rightarrow OH^{\bullet} + OH^- \qquad (2)$$

Hence under $N_2O$ atmosphere, the only reducing species generated in the water phase were the alcohol radicals, which are weaker reducing agents than the solvated electrons. In these conditions, irradiation induced formation of aggregates of spherical particles of 4 nm (see **Figure S8 d**). These experiments suggest that the reduction by strong reducing agents, the solvated electrons, is necessary to obtain the 3D-porous nanostructures. The solvated electrons probably induce seeds of the adequate structure, which further slowly grow by the slow reduction of the adsorbed metal ions by the pentanol radicals, the growth being directed in 1D and 3D because of the preferential adsorption of the CTAB surfactant and the confinement imposed by the mesophase template.

**Nanoballs with tunable composition**

Similar porous nanoballs were obtained with different initial salt compositions Pd/Pt = 1:3, 1:1 and 3:1 for a fixed swelling ratio O/W=1.5. As shown in TEM images (**Figure S5**) nanoballs with equal pore size (16 nm) were obtained for these three compositions. The selected area electron diffraction (SAED) patterns confirmed the cubic structure of the Pd-Pt nanoballs with four main characteristic diffracting rings corresponding respectively to (111), (200), (220) and (311) planes. The measures on SAED patterns were not enough sensible to observe a detectable variation of the distance planes with the nanoballs composition. EDS analysis showed the co-presence of Pd and Pt. The intensities of energy peaks assigned to Pd and Pt show noticeable evolution which follow the same trend as the initial Pd/Pt ratio. Moreover small Br signal, originating from the surfactant CTAB, was detected which indicated its presence even after several washings.

X-ray photoelectron spectroscopy (XPS) characterizations were performed on the Pd-Pt nanoballs of different compositions. **Figure 6a** shows that the nanostructures XPS responses exhibit two peaks at around 335.4 and 340.6 eV specific to the Pd $3d_{5/2}$ and Pd $3d_{3/2}$ spin orbit split core levels.[63] The 3d$_{5/2}$ binding energy contribution at 335.4 eV is associated to a (Pd$^0$) response demonstrating that Pd is present in its reduced form in the nanoballs. The asymmetric shapes of each contribution are always present in a reproducible way, in agreement with the metallic character of Pd in the nanostructures. In the same energy region as Pd 3d, a side signal can be detected in the low binding energy window when Pt quantity increases, which corresponds to the contribution of the Pt 4d$_{3/2}$ core level. Data (**Figure 6b**) show also two other asymmetric peaks centered at around 71.3 and 74.7 eV which are specific to the Pt 4f$_{7/2}$ and Pt 4f$_{5/2}$ core levels of the present Pt atoms and are attributed to metallic platinum (Pt$^0$).[63],[64] A third peak (at 68.8 eV) specific of the Br 3d core level is sometimes observed, whose origin is presumably the surfactant which was not completely removed from the metallic nanostructures. Very good fittings of Pt$_{4f}$ and Pd$_{3d}$ peaks are achieved using only one contribution with the specific metallic asymmetry. This demonstrates the complete reduction of Pt and Pd in the nanoballs (**Figure S9**). XPS atomic percent ratios of Pd and Pt give numerical values close to 0.3, 0.7, 7.5 respectively for initial salt compositions Pd/Pt: 0.33, 1, 3. Because the composition of the outer part of the nanoballs was evaluated from XPS analysis, segregation phenomenon could be detected. The fact that the relative intensities determined by XPS follow roughly the experimental Pd/Pt ratio suggests that any trend for segregation can be discarded. However, to get more information about the fine structure of the nanoballs high-angle annular dark field scanning transmission electron microscopy (HAADF-STEM) and EDS elemental profiles were performed.

HAADF-STEM images of the nanoballs show that they are composed by an alloy of Pd/Pt. **Figure 4b** shows the atomic composition along a tip of an external nanowire composing the nanoball. The contrast is roughly proportional to Z$^2$ where Z is the atomic number. Therefore due to the large difference of atomic number of Pt and Pd the contrast is very easy to see. Every spot in the image of Figure 4b corresponds to an atomic column. The bright spots correspond therefore to column richer in Pt and the lighter spots to columns richer in Pd. We find that the nominal composition of the sample (Pd/Pt = 1:1) is roughly followed in the image. Nevertheless, inside a nanowire small monodomains richer in Pt or in Pd can be observed (**Figure 4b1).** The corresponding FFT image (**Figure 4b2**) indicates an orientation with respect to the electron beam in the <110> direction. A sequence of HAADF-STEM pictures show that Pt atoms can segregate to the surface and move under the electron beam influence (see **Figure S10 b-f**).

Characterization of the Pd and Pt elemental distribution across external nanowires was performed by STEM-EDS line scan techniques. The composition of the nanowires follows the initial Pd/Pt ratios of the precursors (**Figure 7**). The composition is very homogenous along the diameters of the nanowires and along their lengths.

Surface segregation mechanism in alloyed clusters is of great importance for controlling their morphology, composition and properties. The arrangement of the metal atoms in the particle is important in catalysis. Because of its lower surface energy, Pd prefers to segregate to the surface of Pd-Pt clusters.[65-69] Chemical ordering in PdPt nanoparticles has been recently studied via DFT calculations combined with symmetry orbit approach. Multishell patchy arrangements are predicted.[70] These arrangements allow the system to minimize the number of weak Pd-Pd bonds with respect to stronger Pt-Pt and Pt-Pd bonds, which is energetically advantageous as Pd-Pt mixing is exothermic. In Pd-Pt nanoballs, we did not observe Pd segregation at the surface. Nevertheless, HAADF-STEM observations show the nanowires look like a patchwork of small nanodomains richer in Pd or in Pt.

In conclusion, the XPS, HAADF-STEM and EDS experiments show that porous alloyed Pd-Pt nanoballs of controlled composition are synthesized in the swollen hexagonal mesophases.

**CONCLUSION**

Hexagonal mesophases composed of oil-swollen surfactant-stabilized tubes arranged on a triangular lattice in water and doped with metallic salts were used as soft templates for the synthesis of porous bimetallic nanostructures. The bimetallic Pd/Pt nanostructures obtained by a slow and homogeneous reduction provided by radiolysis are porous nanoballs whose alloyed structure was confirmed by using HAADF-STEM, EDS and XPS. The nanoballs comprise 3D-connected nanowires of typical thickness 2.5 nm forming hexagonal cells. We have shown using different techniques (electron microscopy and small-angle X-ray scattering) that the pore size can be varied from 10 to 55 nm in a controlled fashion, and have demonstrated that it is directly determined by the diameter of the oil tube of the doped mesophases, proving thus a direct templating effect of the mesophase. Very few examples have been reported for the preparation of three-dimensional porous bimetallic nanostructures[52, 53] and control over such a large range the porosity of nanomaterials is a premiere.

Our templating approach allows therefore the synthesis of bimetallic nanoballs of tunable three-dimensional porosity and composition. These nanostructures might find applications in catalysis, electrocatalysis and sensing.

**METHODS**

**Materials.** We used tetraaminepalladium(II) dichloride, $Pd(NH_3)_4Cl_2$ (purity 99.99% form Aldrich) and tetraamineplatinum(II) dichloride, $Pt(NH_3)_4Cl_2$ (99% purity from Aldrich) as metallic salts. Propanol ($\geq$ 99% for HPLC), cetyltrimethylammonium bromide CTAB ($\geq$ 98%), cetyltrimethylammonium chloride CTAC ($\geq$ 98%), cetylpyridinium bromide CPBr ($\geq$ 98%), pentanol ($\geq$ 99%), cyclohexane (> 99%) were obtained from Sigma-Aldrich. High purity (impurities less than 1 ppm) nitrogen gas from Air Liquid was used to degas the mesophases. A few experiments have been conducted under $N_2O$ gas (from Air Liquid).

**Mesophases preparation and composition.** The mesophases were prepared as follows: 1.03 g of CTAB (or 0.90 g of CTAC or 0.89 g of CPBr) was dissolved in 2 mL of an aqueous solution containing salts (NaCl and/or metallic salts). After a vigorous agitation and after a few minutes at 50 °C, the surfactant (CTAB, CTAC or CPBr) had completely dissolved to give a transparent and viscous micellar solution. The subsequent addition under stirring of cyclohexane in the micellar solution lead to a white unstable emulsion. The cosurfactant, pentanol, (between 240 and 320 μL) was then added to the mixture, which was strongly vortexed for a few minutes. This lead to a perfectly translucent, birefringent and stable gel: a hexagonal mesophase.

We varied the swelling of the mesophase by changing the volume ratio of oil over water (O/W). Sample compositions are given in **table 1**.

To prepare metallic nanostructures, we doped the mesophases by replacing totally or partially NaCl by metallic salts. Two types of samples were prepared. In all cases, the total metallic salt concentration in the mesophases was fixed at 0.1 M. In a first series of experiments, the metallic salt composition was fixed at Pd/Pt = 1:1, and the O/W ratio and total salt concentration were varied concomitantly following the compositions given in **table 1**. Hence these samples contained always 0.1 M metallic salt and a variable amount of NaCl (from 0 for a swelling ratio O/W of 1.5 to 0.8 M for a swelling ratio of 6.1). In a second series of experiments, samples with O/W=1.5 and metallic salt concentration 0.1 M (no NaCl) were prepared with different compositions Pd/Pt = 1:3, 1:1 and 3:1.

| [NaCl] [M] | O/W (v/v) | $d_c$ [nm] | D [nm] | $e = d_c - D$ [nm] |
|---|---|---|---|---|
| 0.1 | 1.5 | 16.0 | 13.2 | 2.8 |
| 0.2 | 2.0 | 21.5 | 18.6 | 2.9 |
| 0.3 | 2.5 | 24.7 | 22.1 | 2.6 |
| 0.4 | 3.25 | 28.5 | 26.3 | 2.2 |
| 0.5 | 3.75 | 37.9 | 35.4 | 2.5 |
| 0.6 | 4.5 | 40.8 | 38.8 | 2.0 |
| 0.7 | 4.9 | 47.9 | 45.9 | 2.0 |
| 0.8 | 5.5 | 52.0 | 50.3 | 1.7 |
| 0.9 | 6.1 | 59.2 | 57.7 | 1.5 |
| 1 | 6.7 | 61.7 | 60.4 | 1.3 |

**Table 1:** Composition and characteristic sizes, as determined by small-angle X-ray scattering, of the undoped hexagonal mesophases. For all samples, the surfactant over water volume ratio is fixed at 0.51, the NaCl concentration and the oil over water volume ratio (O/W) are varied as indicated in the table. $d_c$ is the lattice parameter, D the diameter of the oil-swollen tubes and $e = d_c - D$ is the thickness of the water channel between adjacent tubes (see a scheme in the inset of **Figure 1a**).

**Radiolytic synthesis of Pd-Pt nanostructures in mesophases**. Mesophases doped with Pd and Pt metallic salts were used to synthesize bimetallic structures as follows. The mesophases were incorporated in glass vessels with a rubber plastic septum, centrifuged at 3000 rpm for 20 minutes and deoxygenated under a $N_2$ flow. The mesophases were then exposed to γ-irradiation at room temperature for 48 hours (irradiation dose of 91.2 kGy) under $N_2$ atmosphere for slow and homogeneous radiolytic reduction. The γ-irradiation source, located at Orsay, was a $^{60}$Co gamma-facility of 7000 Curies with a maximum dose rate of 1900 Gy h$^{-1}$. The samples, which were initially translucent and slightly yellow gels, turned into black gels after irradiation. After reduction, the nanomaterials were extracted in 2-propanol, centrifuged, and washed several times to eliminate the surfactant and the salt.

**X-ray scattering experiments**. Undoped and doped mesophases were analyzed by small-angle X-ray scattering (SAXS) both before and after γ-irradiation. Dry powders of nanomaterials were also analyzed by SAXS and wide-angle X-ray scattering (WAXS). The mesophases were inserted in glass capillaries of 1.5 mm diameter, and the powders were sandwiched between two kapton foils. Experiments were performed with an in-house setup of the *Laboratoire Charles Coulomb, "Réseau X et gamma", Université Montpellier 2, France*. A high brightness low power X-ray tube, coupled with aspheric multilayer optic (GeniX$^{3D}$ from Xenocs) was employed, which delivered an ultralow divergent beam (0.5 mrad). Scatterless slits were used to give a clean 0.8 mm diameter X-ray spot with an estimated flux around 35 Mph/s at the sample position. We worked in a transmission configuration. The scattered intensity was collected on a two-dimensional Schneider 2D image plate detector prototype, at a distance of 1.9 m (for SAXS) and 0.2 m (for WAXS) from the sample. The experimental data were corrected for the background scattering and the sample transmission. Samples of hexagonal mesophases can be unambiguously identified by the occurrence of three Bragg peaks in the scattered intensity, whose positions are in the ratio $1:3^{1/2}:2$. In addition, the characteristic sizes of the mesophases (scheme of a cross-section of a hexagonal phase in the inset of **Figure 1a**) can be uniquely determined by the peak positions and sample composition. The center-to-center distance between adjacent tubes, or lattice parameter, reads $d_c = \dfrac{4\pi}{\sqrt{3}\, q_0}$, where $q_0$, the position of the first peak. One can further relate the lattice parameter to the diameter of the oil-swollen surfactant stabilized tubes, $D$:
$D = 2d_c \left( \dfrac{\sqrt{3}}{2\pi}(1-\phi_p) \right)^{1/2}$, where $\phi_p$ is the volume fraction of polar medium (including the polar heads of the molecules of surfactant and cosurfactant) in the sample. The thickness of the water channel between two adjacent tubes reads simply $e = d_c - D$.

**Transmission electron microscopy observations and analysis.** A drop of the suspension of metallic nanostructures in 2-propanol was evaporated on carbon-coated copper grids for transmission electron microscopy (TEM) observations. TEM experiments were performed on a JEOL JEM 100 CXII transmission electron microscope at an accelerating voltage of 100 kV. For high resolution electron microscopy (HRTEM) images we used a JEOL JEM 2010 equipped with a LaB$_6$ filament and operating at 200 kV. The images were collected with a 4008 X 2672 pixels CCD camera (Gatan Orius SC1000). The chemical analyses were obtained by a selected energy-dispersive X-ray spectroscopy (EDS) microanalyzer (PGT-IMIX PC) mounted on the JEM 2010.

The samples were also characterized using the aberration (Cs) corrected JEOL JEM-ARM 200F: 200 kV FEGSTEM/ TEM equipped with a CEOS Cs corrector on the illumination system.

The probe current used for acquiring the High-angle annular dark field (HAADF)- and the bright field (BF)-STEM images was 9C (23.2 pA); the condenser lens aperture size was 40 µm. HAADF-STEM images were acquired with a camera length of 8 cm/6 cm and the collection angle of 68–280 mrad/90–270 mrad. The BF-STEM images were obtained using a 3 mm/1mm aperture and a collection angle of 17 mrad/5.6 mrad (camera length in this case was 8 cm). The HAADF as well as the BF images were acquired using a digiscan camera. EDS measurements for linescan profiles were obtained with a solid state detector and software for two dimensional mapping from EDAX.

For each type of samples, nanoballs and pores sizes were evaluated by averaging several measurements on TEM images using Image J software.

**X-ray photoelectron spectroscopy experiments.** The X-ray photoelectron spectroscopy (XPS) analysis was performed on In foils. Sample drops were deposited on the foils and dried under $N_2$ flow. The XPS analyzer was a Thermo Electron ESCALAB 220i-XL. Either a non-monochromatic or a monochromatic X-ray Al Kα line was used for excitation. The photoelectrons were detected perpendicularly to the support. A constant analyzer energy mode was used with pass energy of 20 eV. Element atomic percentage determinations were performed using peak area corrected by sensitivity factors using the Thermo Avantage data base. Fitting procedures were also performed using the Thermo Avantage Program.

*Acknowledgements.* The authors thank the *Région Languedoc Roussillon* for their financial support to the « *Réseau X et gamma de l'Université Montpellier 2* ». MJY would like to acknowledge The Welch Foundation Agency Project # AX-1615. "Controlling the Shape and Particles Using Wet Chemistry Methods and Its Application to Synthesis of Hollow Bimetallic Nanostructures". The authors would also like to acknowledge the NSF PREM Grant 0934218 and the NSF Grant 1103730 "Alloys at the Nanoscale; The Case of Nanoparticles Second Phase". The authors thank J.L. Rodriguez Lopez (Instituto Potosino de Investigacion Cientifica y Tecnologica, San Luis Potosi, Mexico) for helpful discussion and Jackie Vigneron (Institut Lavoisier, Université Versailles Saint Quentin-en-Yvelines) for his assistance for XPS experiments.

**Supporting information available:** SAXS signal of the pure NaCl doped mesophases, SAXS signals of the mesophase before and after irradiation, additional TEM observations of Pd-Pt porous nanoballs (of different compositions with the corresponding EDS spectra or with the composition 1:1 and obtained at large swelling), TEM images showing the effects of the nature of the surfactant and of the $N_2O$ atmosphere on the Pd-Pt nanostructures, XPS spectra fittings and additional BF-STEM, HAADF-STEM and HRTEM images of the nanowires forming the nanoballs are available free of charge via the internet at http://pubs.acs.org.

**FIGURES**

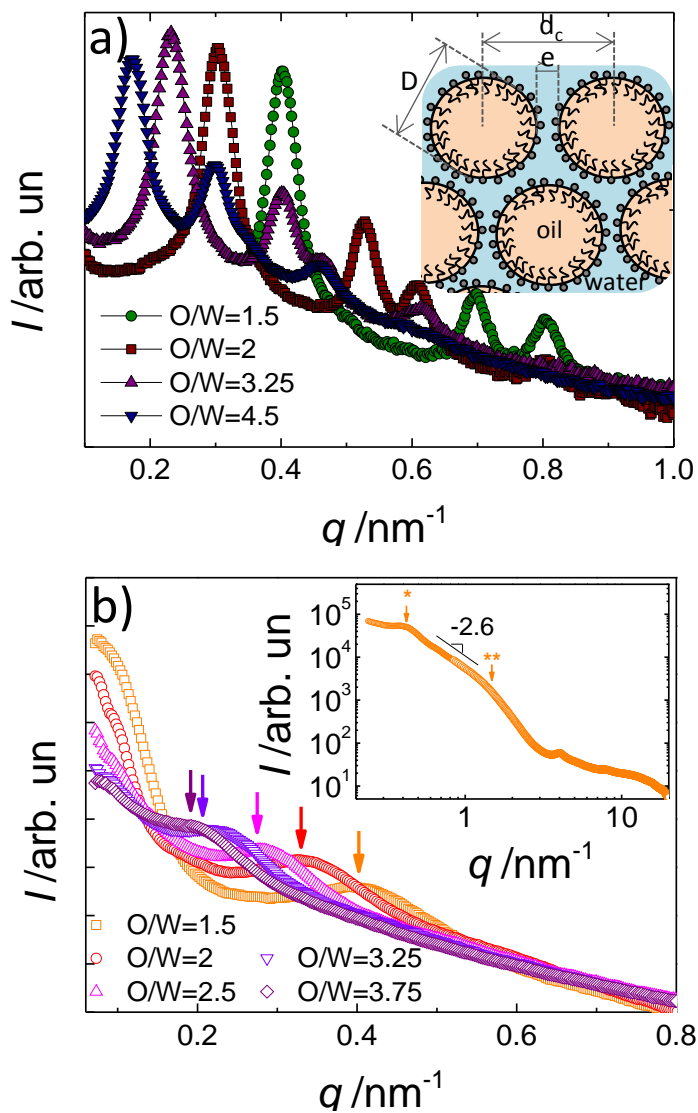

**Figure 1:** Scattered intensity as a function of wave vector $q$ of doped mesophases comprising 0.1 M metallic salts (Pd/Pt = 1:1) **(a)**, and of the Pd-Pt nanoballs synthesized in the mesophases **(b)**. In **(a)** data were taken before γ-irradiation of the doped mesophases with different swelling ratios O/W as indicated in the legend. Inset: scheme of the cross-section of a swollen hexagonal mesophase. In **(b)** data correspond to nanoballs with various pore diameters and prepared from template mesophases with various swelling ratios, O/W, as indicated in the legend. The arrows point to the peak positions. Inset: full scattering curve combining small-angle X-ray scattering and wide-angle X-ray scattering for nanoballs prepared in a mesophase with a swelling ratio O/W = 1.5. The arrows point to the peak position (*) from which the pore diameter is derived and to the departure from the power law regime (**) which provides a characteristic size (4.5 nm) comparable to the thickness of the metallic nanowires.

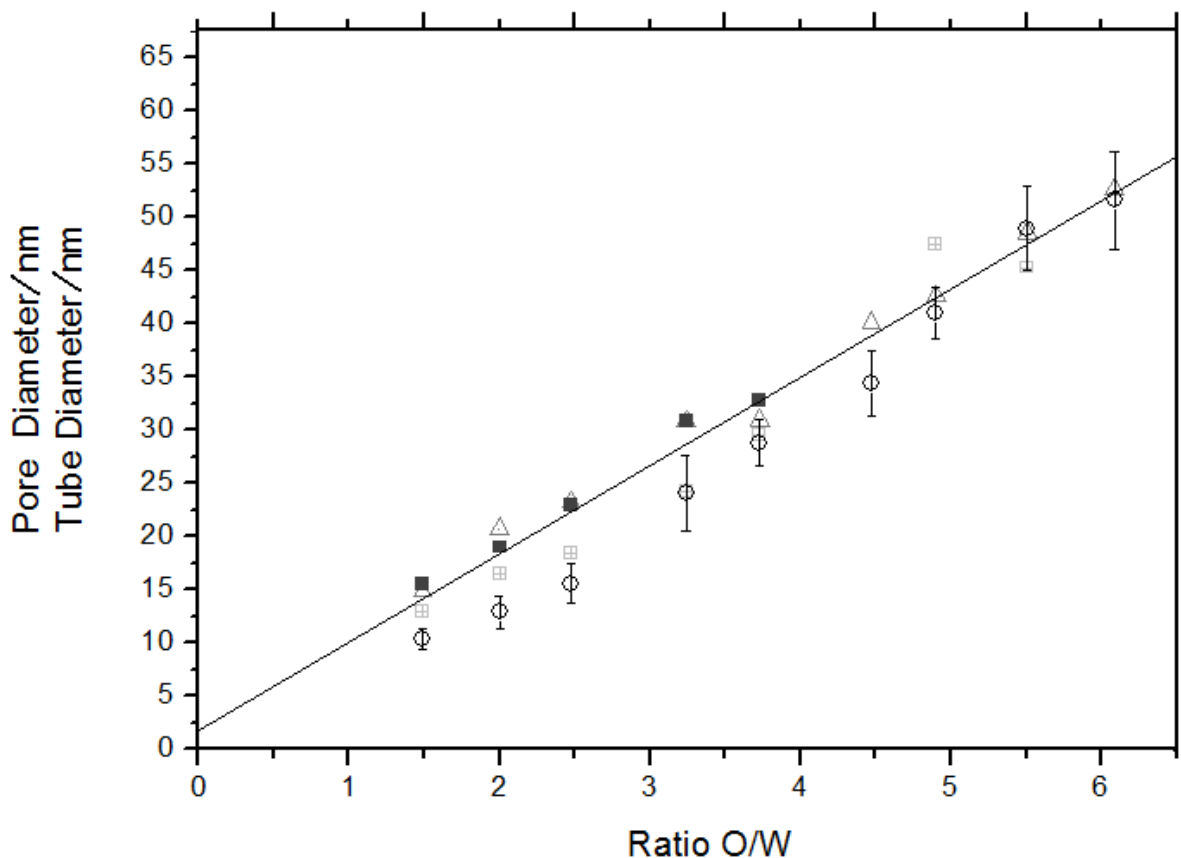

**Figure 2:** Tube diameter of the soft templates and pore size of the nanoballs (0.1 M metallic salts, Pd/Pt = 1:1) as a function of the swelling ratio of the soft templates. The diameter was measured from the SAXS of the soft templates, before (empty dark grey upward triangles) and after irradiation (empty light grey squares), and the pore size was determined by SAXS (full black squares) and by electron microscopy (empty black circles). Symbols are experimental data and the line is a linear fit of the data.

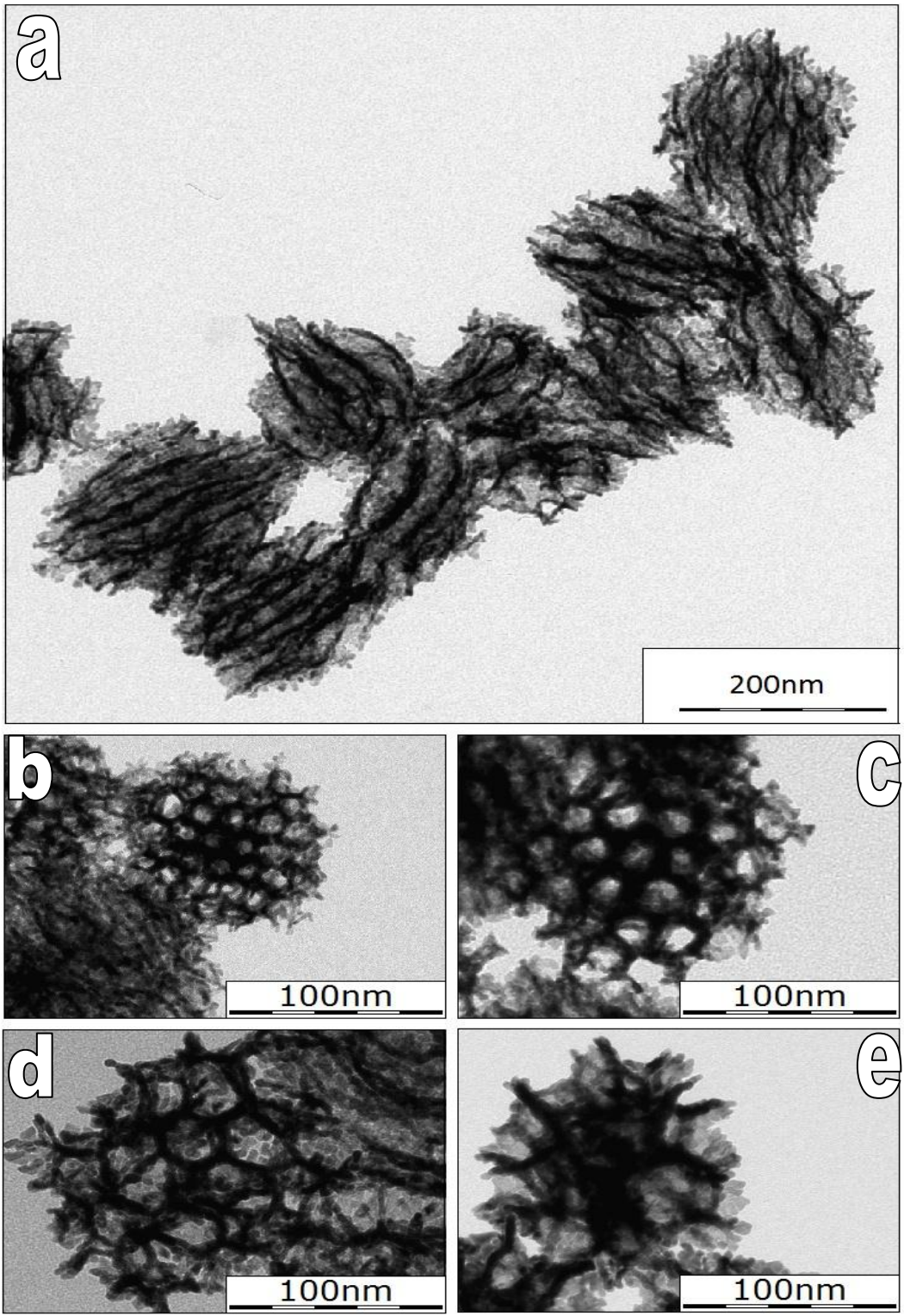

**Figure 3:** Representative TEM images of assembly or single nanoballs (for Pd/Pt = 1:1) with increasing pore sizes corresponding to a ratio O/W equal to: **a)** and **c)** 2, **b)** 1.5, **d)** 3.25 and **e)** 4.5 and with a concentration of metallic salts in the aqueous phase of 0.1 M.

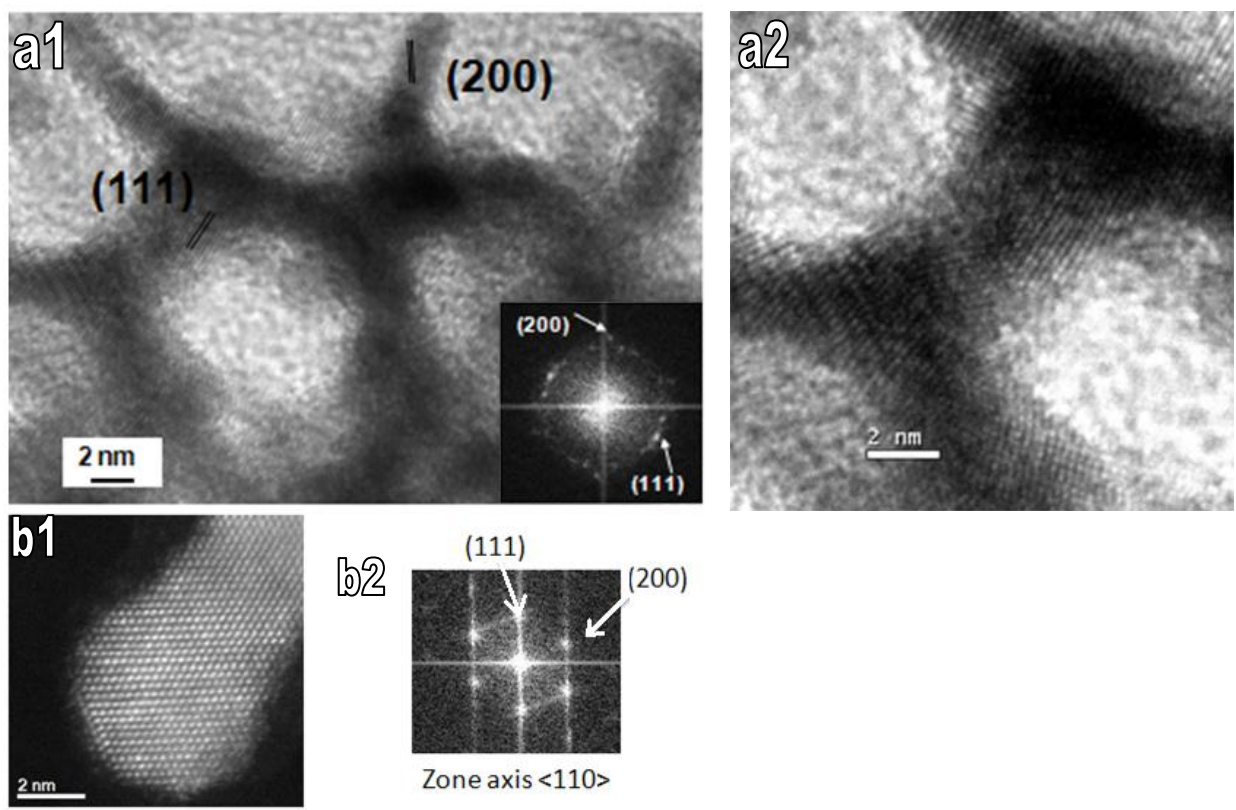

**Figure 4: a1)** High-resolution transmission electron microscopy (HRTEM) image of a hexagonal cell of a nanoball with its corresponding FFT in inset **a2)** highly magnified image of a part of the hexagonal cell, showing nanowire interconnections and **b)** high angle annular dark field scanning transmission electron microscopy (HAADF-STEM) of an external nanowire of a Pd/Pt = 1:1 nanoball **b1)** and its corresponding FFT image **b2)**.

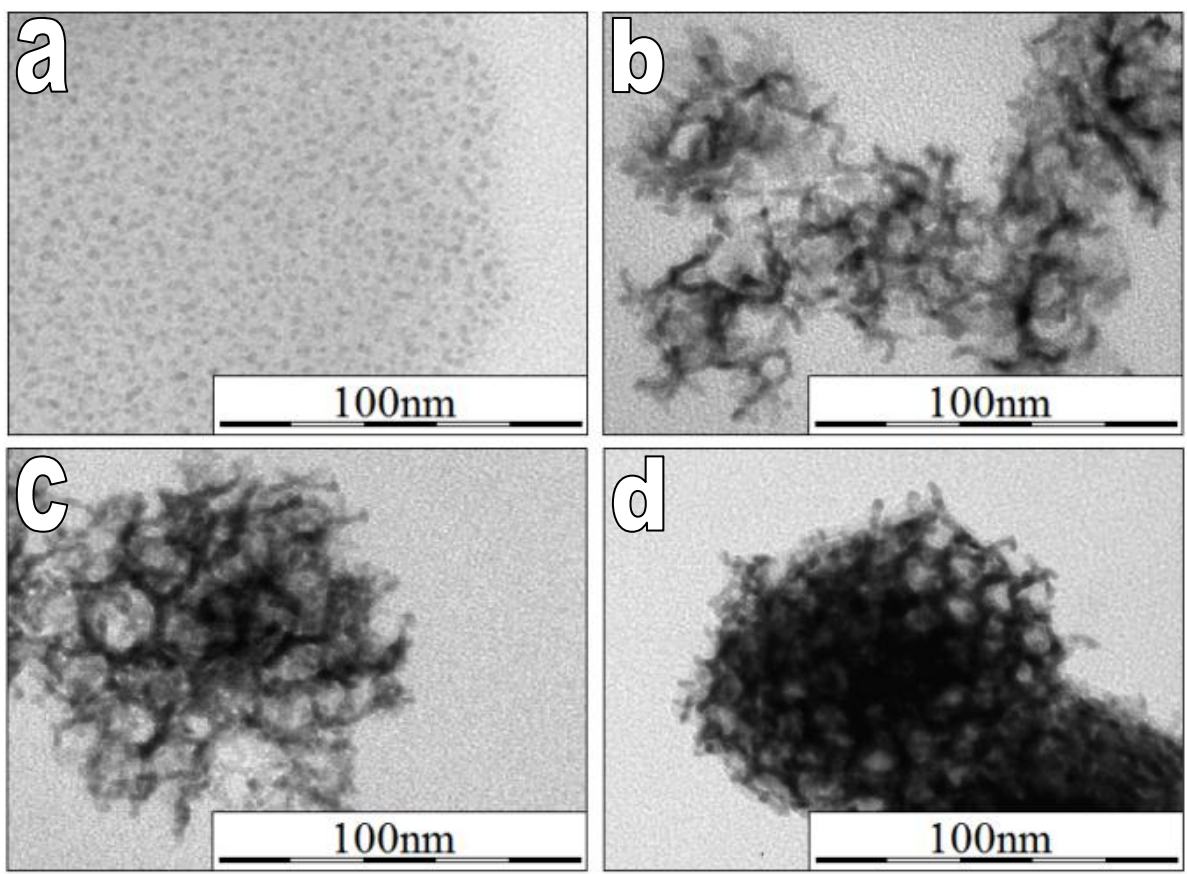

**Figure 5:** TEM images showing the growth of the porous Pd/Pt nanostructures with the dose obtained by irradiation of a doped mesophase with a swelling ration O/W=1.5 and containing 0.1 M of metal salts (Pd/Pt 1:3). The dose rate is 1.85 kGy.h$^{-1}$, and the irradiation time is **(a)** 5h (9.25 kGy), **(b)** 12h (22.2 kGy), **(c)** 24h (44.4 kGy) et **(d)** 48h (88.8 kGy).

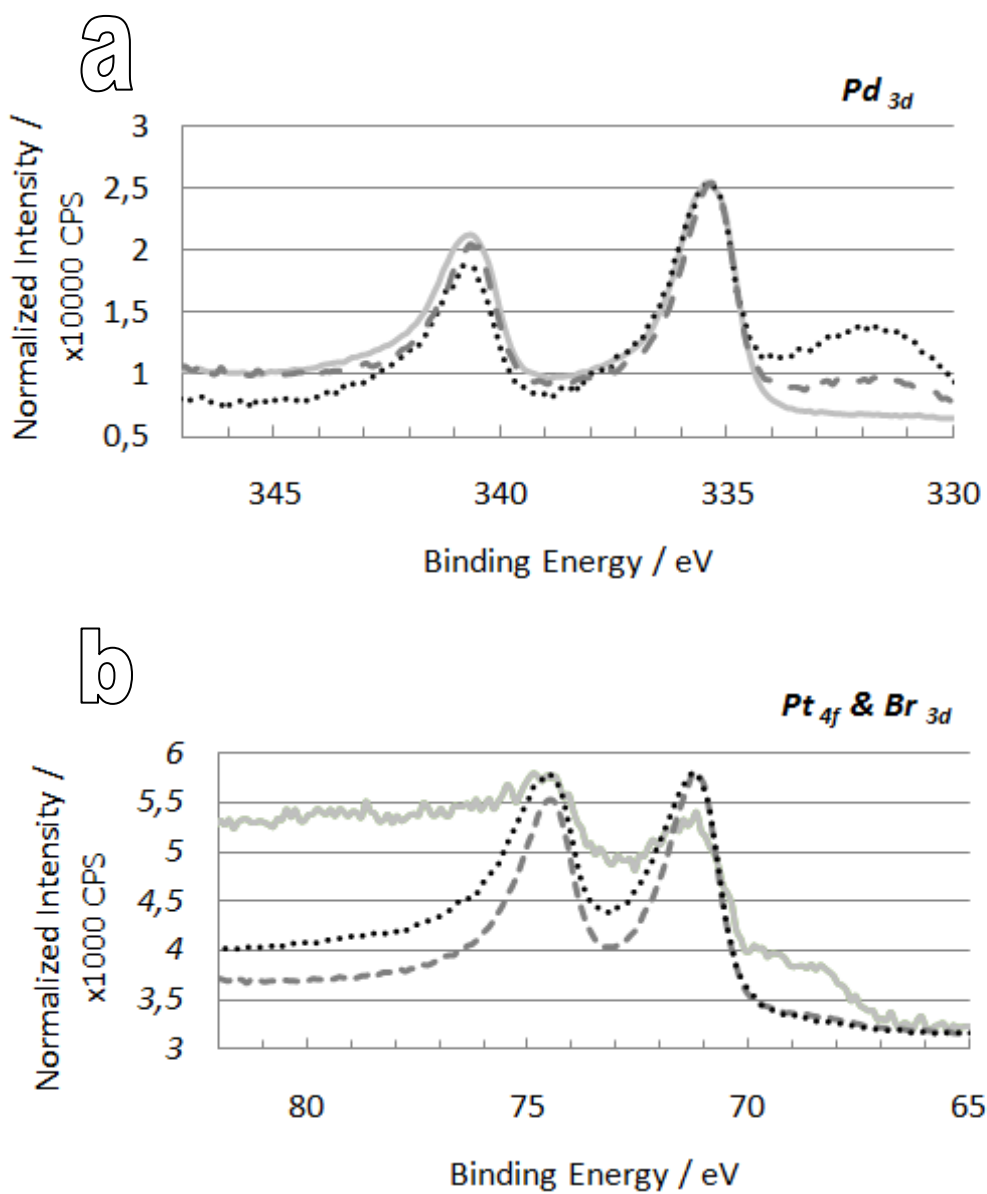

**Figure 6:** X-ray photoelectron spectroscopy spectra of nanoballs of different compositions deposited on InP foils: Pd/Pt = 1:3 in black dots, Pd/Pt = 1:1 in dark grey dashed line, Pd/Pt =3:1 in full light grey line. Samples were all prepared in mesophases with a swelling ratio O/W=1.5.

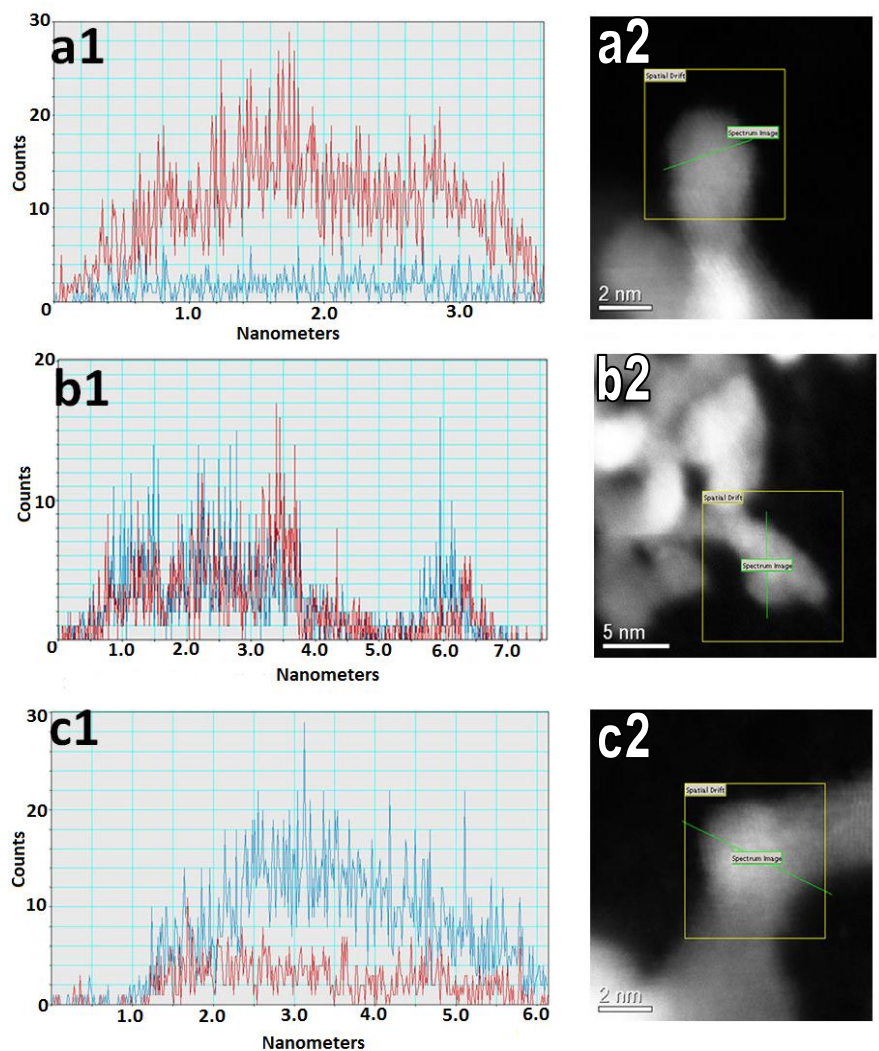

**Figure 7:** Energy dispersive X-ray spectroscopy line scans across external nanowires of nanoballs (profiles were taken along the green line) and corresponding STEM images for the samples: **(a)** Pd/Pt = 1:3, **(b)** Pd/Pt = 1:1 and **(c)** Pd/Pt = 3:1. The blue line corresponds to Pd-L and red one to Pt-L signal.

# SUPPORTING INFORMATION

Supporting information is available from the Wiley Online Library or from the author.

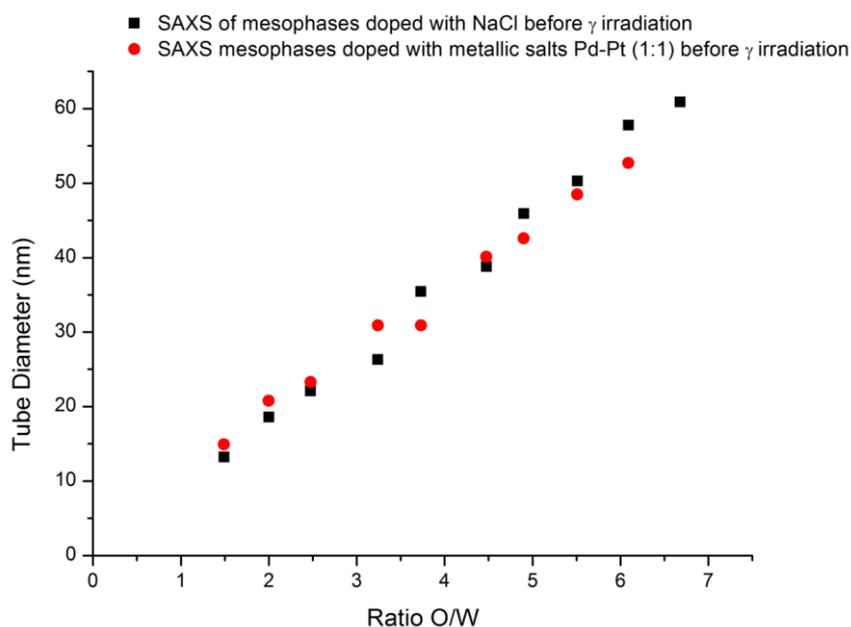

**Figure S1:** Tube diameter of the soft templates doped with NaCl (black squares) or with metallic salts (red points) as a function of the swelling ratio of the soft templates.

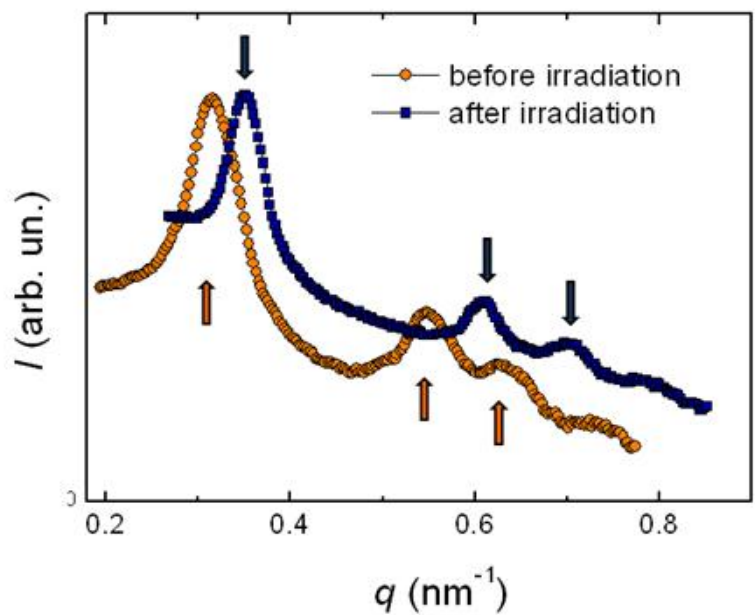

**Figure S2:** Scattered intensity as a function of wave vector $q$ for mesophases comprising 0.1 M metallic salts (Pd/Pt = 1:1), the swelling ratio, O/W, is 2.5. Data before and after γ-irradiation are displayed. The arrows point to the position of the first three Bragg peaks.

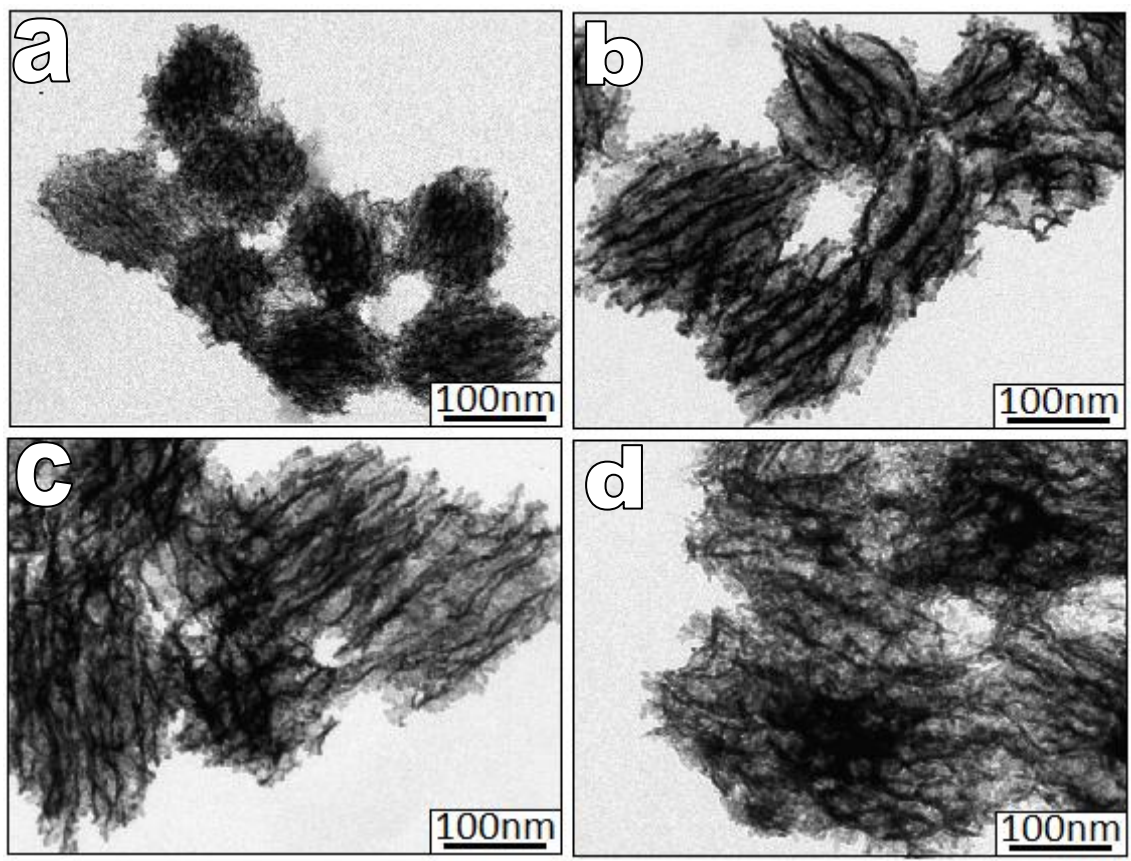

**Figure S3:** TEM images of assembly of Pd-Pt nanoballs (Pd/Pt = 1:1) with increasing pore sizes corresponding to a ratio O/W equal to **a)** 1.5, **b)** 2.5, **c)** 3.75 and **d)** 5.5 and synthesized with a concentration of metallic salts in the aqueous phase of 0.1 M.

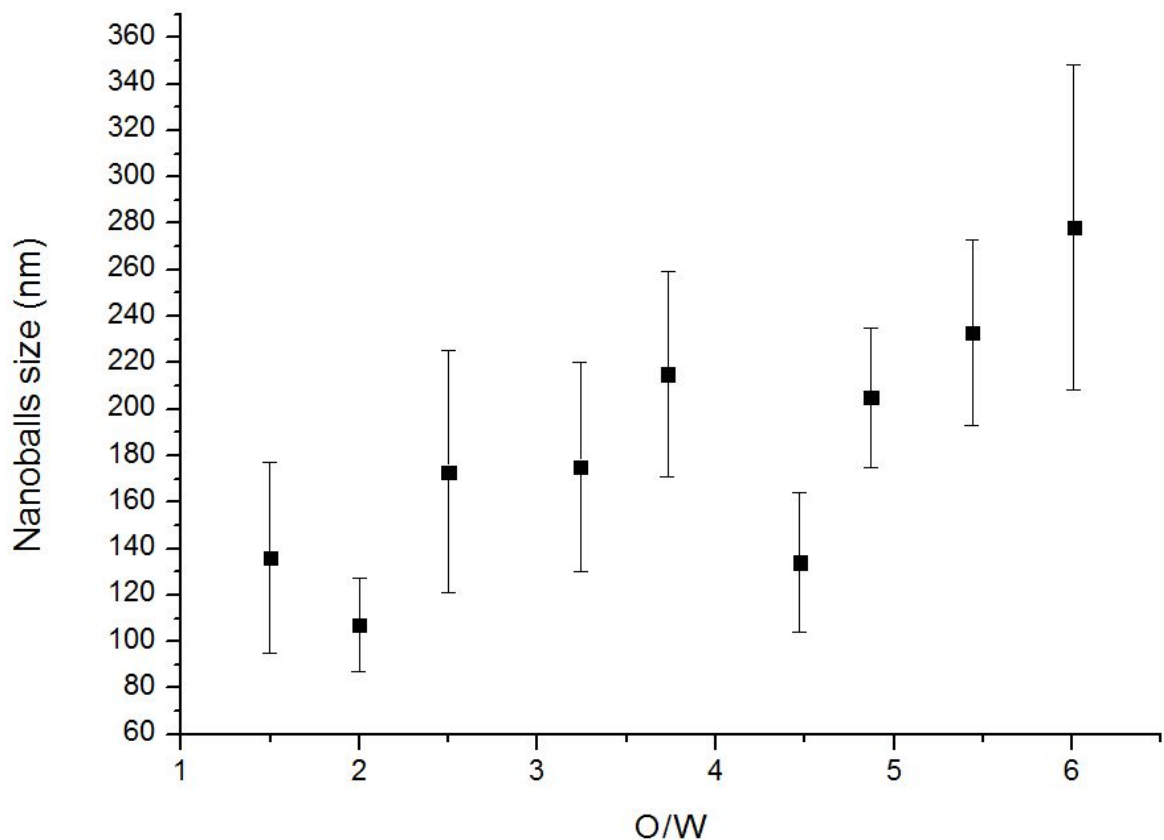

**Figure S4:** Average size of the Pd-Pt nanoballs measured from TEM images *versus* the swelling ratio (O/W).

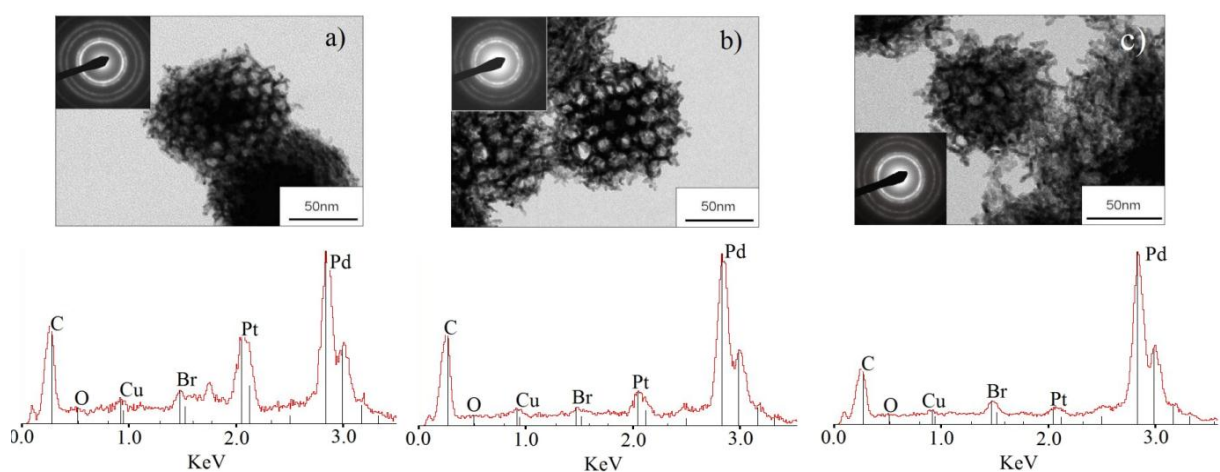

**Figure S5**: TEM images (with corresponding SAED in inset) and the corresponding EDX spectra (acquired in the middle of the nanoballs) of the nanoballs with different composition Pd/Pt **a-** (1:3), **b-** (1:1), **c-** (3:1) prepared in hexagonal mesophases with a swelling ratio O/W of 1.5 and a concentration of metallic salts in the aqueous phase at 0.1 M. A same pore size (16 nm) is measured in a, b and c.

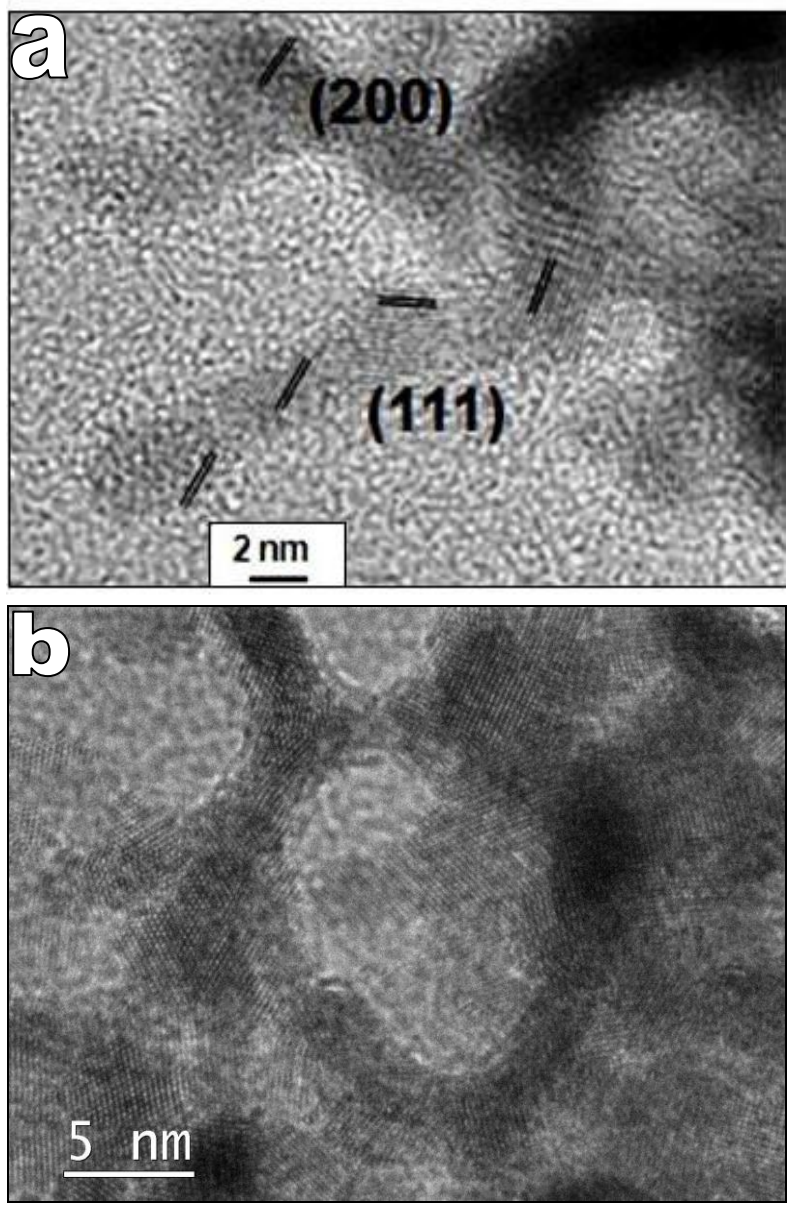

**Figure S6**: High-resolution transmission electron microscopy of **a)** an external nanowire showing the polycrystallinity of the nanoball **b) interconnected nanowires.**

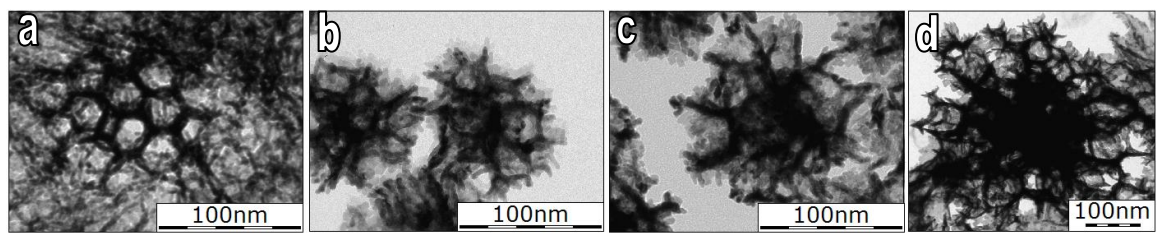

**Figure S7**: Representative TEM images of Pd/Pt = 1:1 nanoballs with increasing pore sizes prepared in hexagonal mesophases with swelling ratio O/W equal to **a)** 2.5, **b)** 3.75, **c)** 4.9 and **d)** 6.1 and with a constant concentration of metallic salts in the aqueous phase of 0.1M.

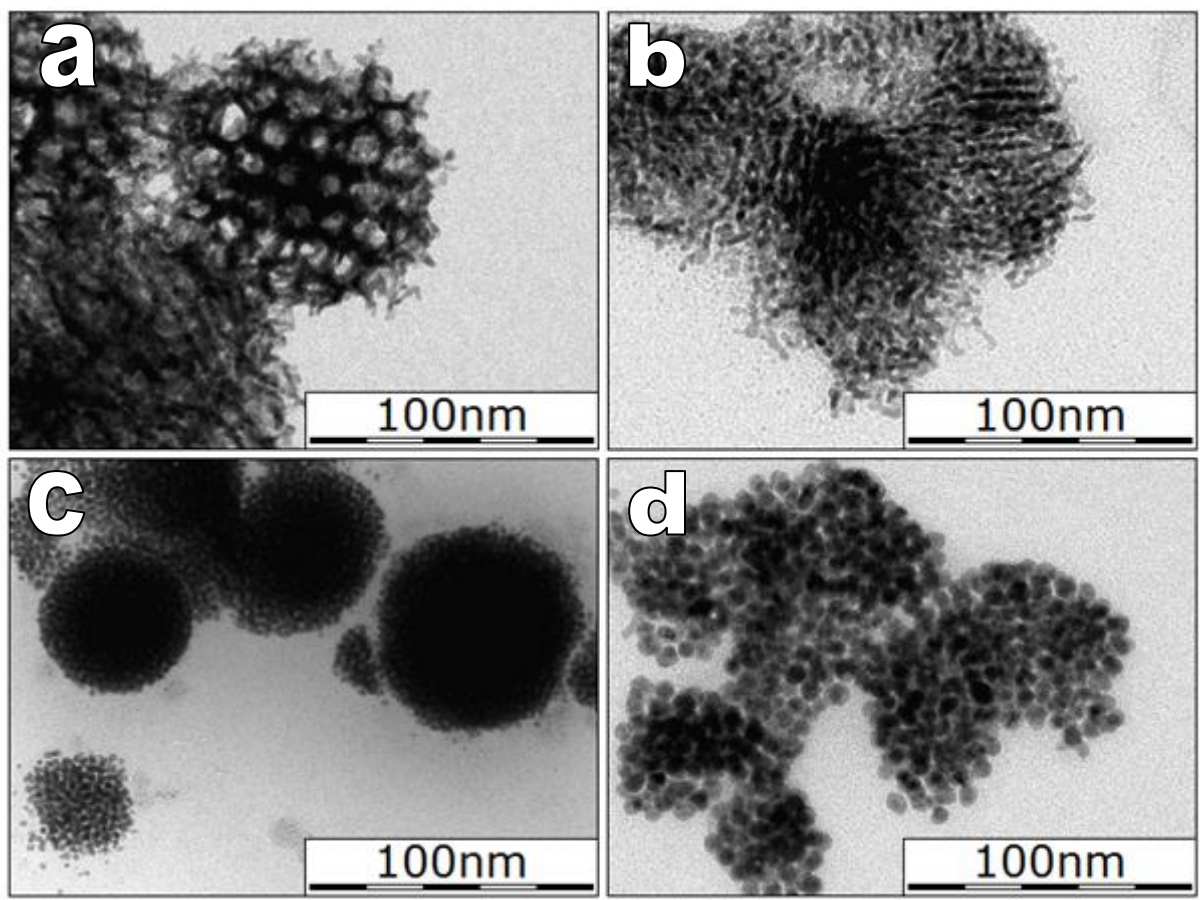

**Figure S8:** TEM images of the Pd-Pt nanostructures (Pd/Pt = 1:1) synthesized in hexagonal mesophases based on different cationic surfactants or different atmosphere during the gamma reduction and containing 0.1 M of metal salts, (O/W=1.5): **(a)** CTAB, **(b)** CTAC and **(c)** CPBr after 48 h irradiation (88.8 kGy) under $N_2$ atmosphere, and **(d)** CTAB under $N_2O$ atmosphere dose rate 1.85 $kGy.h^{-1}$.

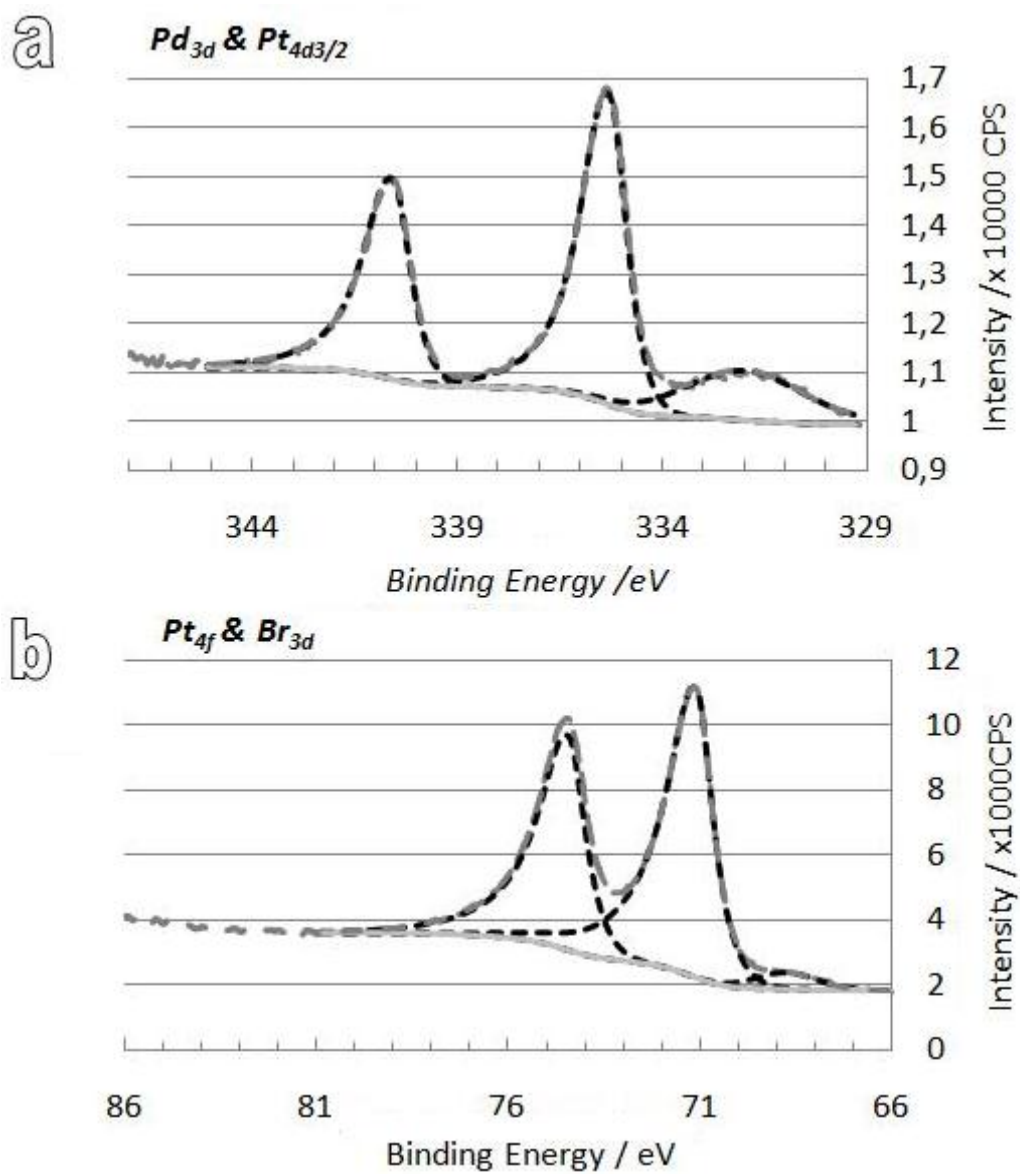

**Figure S9:** X-ray photoelectron spectroscopy spectra of nanoballs obtained with the initial ratio Pd/Pt = 1:1 and deposited on InP foils (the samples were prepared in a mesophase with a swelling ratio O/W=1.5). **(Top)** $Pt_{4f}$ fitting. The low binding energy contribution is attributed to a residue of $Br_{3d}$ signal originating from the precursor. **(Bottom)** $Pd_{3d}$ fitting. The low binding energy contribution is attributed to $Pt_{4d}3/2$ level.

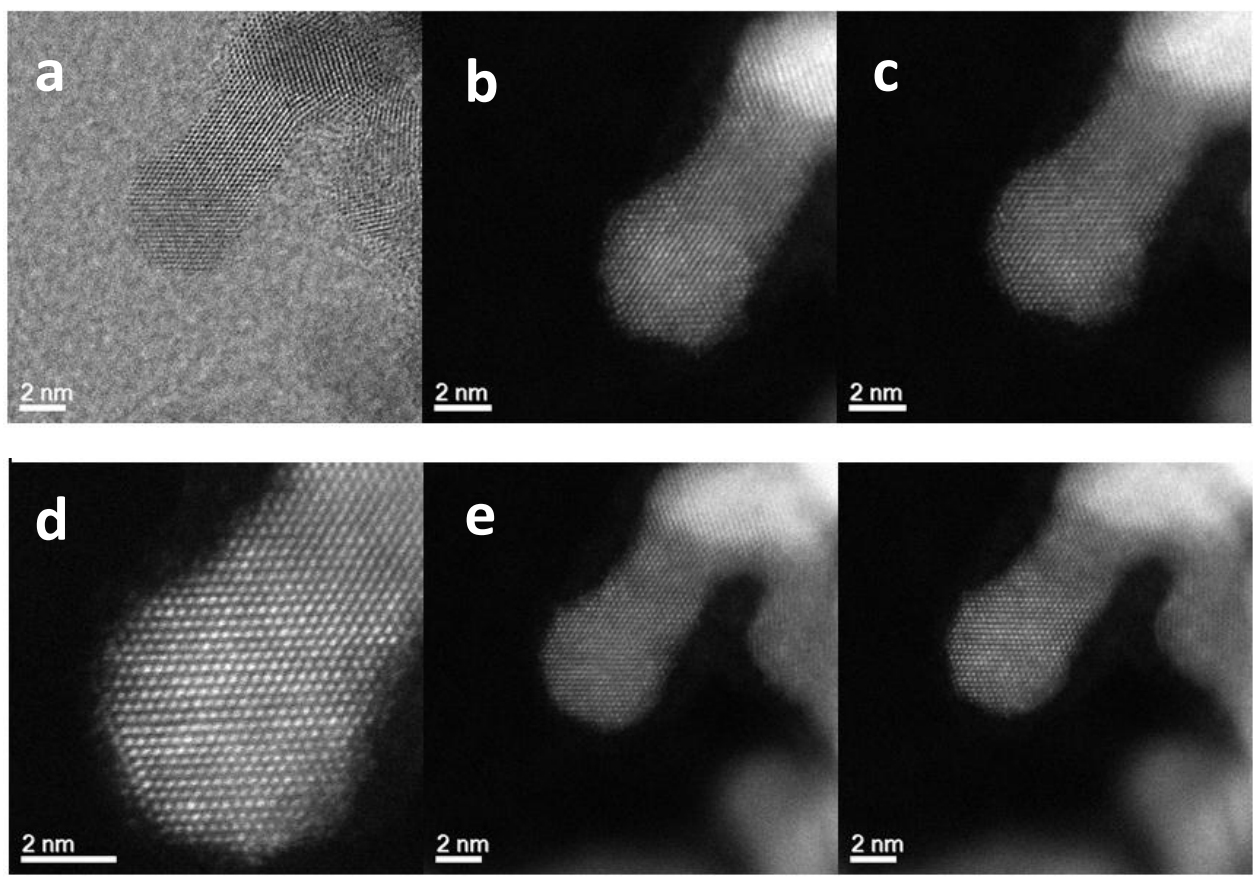

**Figure S10**: Bright field scanning transmission electron microscopy (**a**) and a sequence of high angle annular dark field scanning transmission electron microscopy pictures (**b-f**) of a nanoball prepared in a mesophase with a swelling ration 0/W=1.5 and with a salt composition Pd/Pt 1:1. In one of the tips of the sample: it can be seen that the Pt atoms (bright spots) segregate to the surface and move under the electron beam influence.